\documentclass{WileyMSP-template}

\newcolumntype{C}{>{\centering\arraybackslash \hsize=1\hsize}X}
\DeclareAcronym{arx}{
	short = ARX,
	long = auto-regressive models with exogeneous inputs
}

\DeclareAcronym{lasso}{
	short = Lasso,
	long = least absolute shrinkage and selection operator
}

\DeclareAcronym{nwp}{
	short = NWP,
	long = numerical weather prediction
}

\DeclareAcronym{lce}{
	short = LCE,
	long = linear cloud edge
}

\DeclareAcronym{mcp}{
	short = MCP,
	long = most correlated pair
}

\DeclareAcronym{ccm}{
	short = CCM,
	long = cross-correlation method
}

\DeclareAcronym{clstm}{
	short = Conv-LSTM,
	long = 	convolutional long-short term memory
}

\DeclareAcronym{cmae}{
	short = CMAE,
	long = 	cumulative mean absolute error
}

\DeclareAcronym{mae}{
	short = MAE,
	long = 	mean absolute error
}

\DeclareAcronym{csa}{
	short = CSA,
	long = 	cross-spectral analysis
}

\DeclareAcronym{arima}{
	short = ARIMA,
	long = 	auto-regressive integrated moving average
}

\DeclareAcronym{pv}{
	short = PV,
	long = 	photovoltaic
}

\DeclareAcronym{cmv}{
	short = CMV,
	long = 	cloud motion vector
}

\DeclareAcronym{ghi}{
	short = GHI,
	long = global horizontal irradiance
}

\DeclareAcronym{dni}{
	short = DNI,
	long = direct normal irradiance
}

\DeclareAcronym{dhi}{
	short = DHI,
	long = diffuse horizontal irradiance
}

\begin{document}

\pagestyle{fancy}

\rhead{Submitted to Solar RRL on June 28, 2024.}

\title{How connected cars could capture cloud dynamics - first evidence from two simulation scenarios}

\maketitle


\author{Tobias Veihelmann*, Philipp Reitz, Maximilian Lübke, Norman Franchi}


\dedication{}

\begin{affiliations}
Tobias Veihelmann \\
Chair of Electrical Smart City Systems\\
Friedrich-Alexander-Universität Erlangen-Nürnberg\\
Cauerstr. 7, 91058 Erlangen\\
tobi.veihelmann@fau.de

\end{affiliations}



\begin{abstract}

The rapidly increasing share of fluctuating electricity from photovoltaics calls for accurate approaches to estimate cloud motion, the primary source for the varying power supply. While local sensor networks are prominent in targeting forecast horizons too short for image-based methods, they have minimal spatial coverage. This work presents the first step towards expanding those approaches to spatially scalable sensor networks: With the motivation of using automotive light sensors as a sensor network, two excerpts from a microscopic traffic simulation serve as simulative sensor networks. A fractal-based cloud shadow pattern passes the sensor network areas with defined velocities and directions, which shall be estimated using the cumulative mean absolute error method. The evaluation results indicate that the more extensive observation areas compensate for the dynamics in the sensor network when compared to a reference work with a static sensor grid. Furthermore, this work shows how the estimates deteriorate with lower vehicle penetration rates (PR) and longer building shadows due to a lower solar elevation angle. At a penetration rate of \qty{40}{\percent}, the root mean square errors for both sensor networks are still below \qty{5}{\meter\per\second}. In conclusion, the spatio-temporal characteristics of a vehicle network offer some potential for estimating cloud movements.

\end{abstract}


\section{Introduction}

Photovoltaic power has become a widely used energy source over the past years due to its decreasing cost (\cite{IRENA.2023b,Sens.2022}) and the need for environmentally friendly electricity production. From 2013 to 2022, the worldwide installed photovoltaic (PV) capacity rose steeply from 137\,GW to 1055\,GW \cite{IRENA.2023}. Following the estimates from reputable institutes and the worldwide efforts to mitigate climate change, this is the start rather than the end of the rise of solar energy. According to the International Renewable Energy Agency (IRENA), meeting the \qty{1.5}{\celsius} goal to limit climate change would require another 18,085\,GW increase by 2050 (own calculation, derived from \cite{IRENA.2023b,IRENA.2023}). 

However, there is a central challenge associated with the skyrocketing adoption of PV installations: Like other renewable energy sources such as wind power, PV power fluctuates due to natural influences. These fluctuations cause challenges for the integration into the electric grid \cite{Mbungu.2020,Martinot.2016,Behabtu.2020}, requiring efficient frequency regulation (e.g., \cite{Mbungu.2020}) and absence of grid congestions (e.g., \cite{Nouri.2022,Koster.2023}). A significant research branch thus focuses on developing more and more sophisticated energy storage systems (ESS) to smooth the fluctuations caused by variable renewable energy (VRE) \cite{Behabtu.2020}. Nonetheless, deploying storage systems in sufficient scales to cope with the extreme growth of PV production would require enormous resources that, in addition, are very hard to predict \cite{Yang.2024a}. 

In parallel, a vibrant research branch on solar forecasting has evolved, offering solutions to facilitate grid integration of PV energy \cite{Yang.2024a,Fabel.2024}. Different spatial and temporal requirements demanded diverse approaches to solar forecasting, with the most prominent being sky cameras, satellite-based methods, and numerical weather prediction (NWP) models \cite{Yang.2022}. In addition, there have been attempts to grasp the spatio-temporal dynamics of solar irradiance with distributed sensors \cite{BenavidesCesar.2022}. Admittedly, those sensors usually offer minimal spatial resolution. The widespread distribution of pyranometers from weather stations or solar panels prevents the capture of cloud dynamics. Thus, large-scale sensor networks with reduced accuracy are limited to statistical methods (e.g., \cite{BenavidesCesar.2022}). 

In contrast, microscale local sensor networks can detect cloud dynamics. Specific sensor network layouts (e.g., \cite{Bosch.2013}, \cite{Chen.2019}, \cite{EspinosaGavira.2022}) enable the estimation of cloud velocities and directions after identifying sensor pairs with the most correlated measurements. These approaches require pyranometers offering very high sampling rates, so most studies employ photodiode pyranometers (e.g., \cite{Bosch.2013,EspinosaGavira.2022,Kuhn.2018}). The problem with these local sensor networks is that they can only capture cloud dynamics within the tiny spatial scale of the network \cite{Yang.2022}. Consequently, their application is bound to forecast horizons below two minutes and areas below a square kilometer \cite{Chu.2021}. 

Hence, extending these small-scale local sensor networks with their ability to capture cloud (shadow) dynamics, high update rate, and high temporal resolution to a significantly more expansive scale would be a great benefit. This work presents a scalable approach utilizing existing sensor nodes designed for another purpose. Thus, no deployment of sensors is necessary. We provide first simulation-based insights on some essential theoretical considerations and limitations that accompany the approach: Many modern cars have wireless communication capabilities and a light sensor attached. The light sensor is necessary, e.g., for dawn detection to automatically switch the light on and off, to adjust the brightness of head-up displays, or to optimize air condition. Existing studies suggest that light sensor measurements targeting illuminance correlate pretty well with a pyranometer’s irradiance measurements \cite{Michael.2020}. Also, Lorenzo et al. \cite{Lorenzo.2014} deployed cheap and straightforward photodiodes in (irradiance) sensor networks to replace pyranometers. Alongside initial successes in using a smartphone's light sensor as a replacement for a pyrheolimeter \cite{DiLaccio.2023}, this research fuels the idea that there may be a way to support solar forecasting with automotive light sensors. This work, however, does not investigate how much automotive light sensors’ measurements resemble irradiance values. Instead, for the following investigations, we assume that car light sensors could perfectly sense solar irradiance and give a first assessment on the following questions: Would sensor networks with perfect irradiance sensors moving on streets be able to capture the dynamics of moving cloud shadows?

Since 2013, more than a handful of papers on local sensor networks have been published (see Table~\ref{tab:lsn}), aiming to detect local cloud motion \cite{Yang.2024f}. These works have proposed various sensor layouts and different approaches to derive cloud motion vectors (CMVs), i.e., cloud movement direction and velocity, from measured irradiance time series from those sensors. Table~\ref{tab:lsn} provides an overview of existing works on local sensor networks with attempts at cloud motion estimation. The table only presents works that deployed the network (in real or simulation) and included a method to capture cloud dynamics.

\begin{table}[htb!]
    \caption{Local sensor networks}
    \label{tab:lsn}
\begin{tabularx}{\textwidth}{X *{5}{C}}
	\toprule
	Study & Network size & Number of nodes & Sampling period & Sensor type & CMV detection \\
	\midrule 
	Bosch et al. (2013) \cite{Bosch.2013} & $<$\,\qty{70}{\square\metre} & 8 & \qty{50}{\milli\second} & pyranometer & \acs{lce}, \acs{mcp} \\
	Bosch et al. (2013)\cite{Bosch.2013b} & $<$\,\qty{4}{\square\kilo\metre} & 96 & \qty{1}{\second} & inverter & \acs{lce}, \acs{ccm} \\
	Lipperheide et al. (2015) \cite{Lipperheide.2015} & \qty{1}{\square\kilo\metre} & 70 & \qty{1}{\second} & inverter & LCE \\
    Wang et al. (2016) \cite{Wang.2016} & $<$\,\qty{1}{\square\metre}& 9 &  \qty{1.5}{\milli\second} & pyranometer & \acs{lce}-curve fitting \\
	Kuhn et al. (2018) \cite{Kuhn.2018} & $<$\,\qty{1}{\square\metre} & 9 & \qty{1.5}{\milli\second} & pyranometer & \acs{lce}-curve fitting\\
	Jamaly and Kleissl (2018)\cite{Jamaly.2018} & \qty{5.8}{\square\kilo\metre} & 9\,216 & \qty{10}{\second} & simulation & \acs{ccm}, \acs{csa} \\
	Jamaly and Kleissl (2018)\cite{Jamaly.2018} & \qty{6.5}{\square\kilo\metre} & 16\,384  & \qty{10}{\second} & simulation & \acs{ccm}, \acs{csa} \\
	Espinosa-Gavira et al. (2020) \cite{EspinosaGavira.2020} & $\leq$ \qty{0.01}{\square\kilo\meter} & 9 to 100 & \qty{0.3}{\second} to \qty{3.3}{\second} & simulation & \acs{lce}, \acs{mcp}, \acs{cmae} \\
    \textbf{This work} & \qty{0.6}{\square\kilo\meter} to \qty{4}{\square\kilo\meter} & 7 to 122 & \qty{1}{\second} to \qty{2}{\second} & simulation & CMAE \\
	\bottomrule
 \multicolumn{6}{p{\linewidth}}{Abbreviations have the following meanings: LCE - linear cloud edge; MCP - most-correlated pair; CCM - cross-correlation method; CSA - cross-spectral analysis; CMAE - cumulative mean absolute error.}
\end{tabularx}
\end{table}

\subsection{Methods to estimate cloud vector motion vectors}

Earlier methods of estimating cloud motion vectors originate from analyzing subsequent satellite or sky images. Clouds can be detected on satellite images since they reflect more light than they let pass, causing their pixels to appear brighter \cite{Aicardi.2022}. Subsequently, with the assumption that image pixel changes are only caused by cloud movement (not dissipation or formation), vectors are selected such that mean square pixel differences for a region around the vector are minimized \cite{Lorenz.2004}. In addition, correlation-based methods have successfully found displacement vectors: A search radius is drawn around a rectangular region of interest. Cross-correlating this region with another image enables the identification of the most-correlated corresponding region in the subsequent image and the resulting displacement \cite{Hamill.1993}. 

For the analysis of ground-based sensor measurements, two of the earliest and most prominent methods to capture cloud shadow dynamics are the \acf{lce} and the \acf{mcp} method, introduced by Bosch et al. \cite{Bosch.2013}. The \acs{lce} method requires a triplet of irradiance sensors. The known angles between the closely aligned sensors and the assumption of a cloud's edge being linear allow us to calculate the cloud motion speed and direction. Also, the \acs{mcp} method requires closely aligned sensors with a specific (or, at least, precisely known) sensor layout. The basic layout presented in \cite{Bosch.2013} is a half-circle of 7 sensors around one central sensor. Once a significant variation in measured irradiance is detected, the measured time series of each half-circle sensor is cross-correlated against the central sensor. The angle of the sensor pair with the highest cross-correlation is assumed to be the cloud movement direction, and the lag value from the cross-correlation enables the velocity calculation \cite{Bosch.2013}. Wang et al. \cite{Wang.2016} employed a sensor layout similar to the one from \cite{Bosch.2013}'s \acs{mcp} layout but with the so-called \acs{lce}-curve fitting method: With the assumption of a linear cloud edge, the time lags for \acs{mcp}-like sensor pairs can be plotted for their directions. As a result, the plotted values will approximate a cosine function. Subsequently, the maximum time shift in the cosine function should represent the time shift for the cloud movement direction \cite{Wang.2016}.

\subsection{Real-world experiments and simulations}

Bosch et al. \cite{Bosch.2013} deployed a pyranometer sensor network and captured real-world measurements for their \acs{lce} and \acs{mcp} methods. They validated the \acs{lce} method with the \acs{mcp} method as the assumed ground truth and found an agreement of $R^2 = 0.977$ for the estimated directions and an agreement of $R^2 = 0.875$ for the velocity estimates. However, the accuracy of the \acs{mcp} method remained uncertain because the true velocities and directions of the cloud shadows were not known. This is a general problem with real sensor networks for the evaluation of methods for capturing cloud dynamics. The theoretical boundaries of such methods have thus been evaluated with simulation-based approaches where the "true" cloud motion vector is modeled and provides a valid ground truth.

Two setups exist to simulate cloud dynamics to create ground truths for validating sensor network-based approaches to detect cloud motion vectors. Jamaly and Kleissl \cite{Jamaly.2018} relied on Large-Eddy Simulation (LES), while other works employed fractal cloud shadow models (e.g., \cite{EspinosaGavira.2020}, \cite{Cai.2013}). LES is based on the Reynolds-averaged Navier-Stokes equations. It enables the calculation of turbulent flows and thus the modeling of atmospheric phenomena, including cloud formation, dissipation and dynamics (e.g. \cite{Jamaly.2018}). Subsequently, irradiance maps can be derived to apply cross-correlation (CCM) and similar methods for CMV estimation. 

Jamaly and Kleissl \cite{Jamaly.2018} applied cross-spectral analysis (CSA) and the cross-correlation method to two LES-based irradiance datasets. They consider each point with LES output a grid point, e.g., $96 \times 96$ grid points for $\qty{2.4}{\kilo\meter} \times \qty{2.4}{\kilo\meter}$, corresponding with a resolution of \qty{25}{\meter}. Removing grid points with low variability in irradiance reduces the sites to 361. They found the cross-correlation method to outperform the cross-spectral analysis.

Espinosa-Gavira et al. \cite{EspinosaGavira.2020} also investigated estimating cloud dynamics with sensors aligned in a grid. In contrast with \cite{Jamaly.2018}, they employed a fractal-based cloud shadow pattern, as introduced by \cite{Beyer.1994} and further developed by \cite{Cai.2013} and \cite{Lohmann.2017}. In this approach, a modified diamond-square algorithm produces a fractal surface. Then, a threshold value is set to create some boundary, with values above the threshold cloudy and below the shadow clear. A transition zone around the threshold allows for more realistic cloud edges, as described in \cite{Lohmann.2017}. Last, the cloud shadow pixels are transformed into a clear-sky index or irradiance map.

In their paper, Espinosa-Gavira et al. \cite{EspinosaGavira.2020} let their irradiance map move over different grid configurations, ranging from \qty{10}{\meter} to \qty{100}{\meter} side length and 3 to 10 grid points (and sensors) per side. They proposed a new method, the cumulative mean absolute error (CMAE), to determine cloud velocity and movement direction. Comparing this method to the MCP and LCE methods showed their approach to outperform those. The results were better with higher resolution grids, with the best results of \qty{1.4}{\meter\per\second} RMSE over 100 simulations. These results were obtained at different speeds and directions with a grid of about 30 meters side length and a total of 100 sensors. 

The validity of approaches requiring a grid of sensors also generalizes to cases where the (simulated) sensors are randomly distributed and not located at the grid point positions. \cite{EspinosaGavira.2020} proposed to employ Kriging to interpolate the values from the distributed sensors to grid points and then apply the CMAE method without any modifications. This approach provided pretty accurate results in a simulation experiment, demonstrating its applicability to non-gridded sensor networks. 

\subsection{Contribution}

Our experiment directly bridges Espinosa-Gavira et al.'s approach \cite{EspinosaGavira.2020} to the dynamic street-network-based sensor network proposed in this work: If, for any time step, measurements from distributed sensor locations are interpolated to a fixed grid, the approach can be applied to our sensor network without any changes. This last building block completes the system model: We employ a microscopic traffic simulation \cite{Veihelmann.2024} to obtain a realistic network of vehicles representing moving irradiance sensors. The fractal cloud model from \cite{Lohmann.2017} contributes a realistic clear-sky index map. Next, we choose two areas from the traffic simulation as observation areas and let the clear-sky index map move over the observation areas for 100 different velocities and directions. For an observation area in a city center with many building shadows to be expected, we model these shadows with the open-source software Blender. Finally, we employ the CMAE method to produce CMVs and compare those to the simulated ground truth.

That way, we want to address the following objectives:
\begin{enumerate}

\item Determine spatio-temporal parameters, i.e., time steps and grid point distances, which enable an accurate agreement between estimated and modeled cloud shadow velocities and directions. We study two different scenarios, one with and one without building shadows. 

\item Investigate how, with good parameterization, the estimation errors change with lower shares of available vehicles.

\item Analyze for the scenario with building shadows how the RMSEs change for two additional times of day with higher traffic volumes and wider building shadows due to lower elevation angles of the sun.

\end{enumerate}

\section{Experimental Section}

\begin{figure}[htb!]
    \centering
    \includegraphics[width=0.9\linewidth]{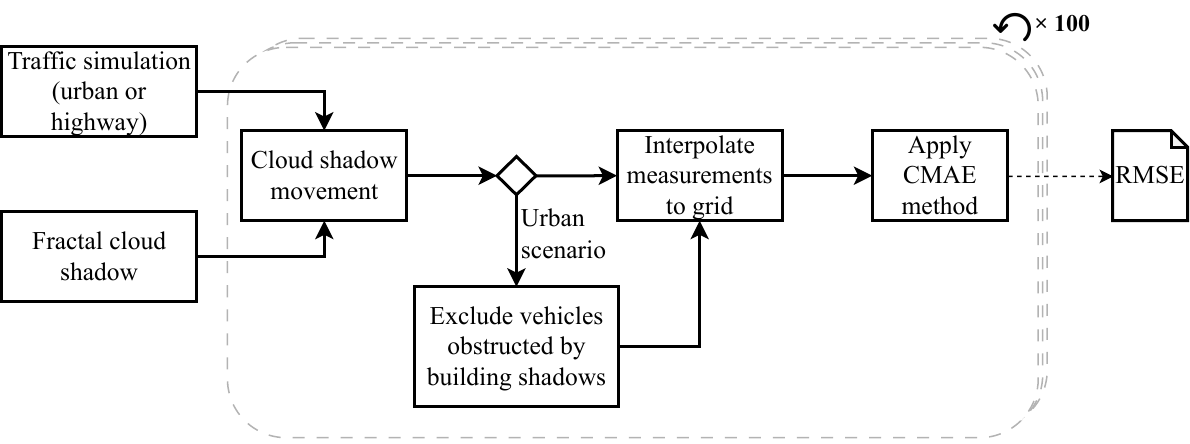}
     \caption{The process steps explained in the experimental section.}
    \label{fig:drawio}
\end{figure}

This section explains the simulation building blocks underlying the analyses in this paper as shown in Figure~\ref{fig:drawio}. First, we present the two traffic simulation-based sensor networks in \ref{section:scenarios} and their interaction with the cloud shadow model in \ref{section:csim} and \ref{section:transition}. Second, we describe a method to determine street segments or measurements which are obstructed by building shadows to exclude the respective measurements from the analysis in \ref{section:shadowmethods}. The section continues with our implementation of the CMAE method based on \cite{EspinosaGavira.2020} and how the measurements are interpolated to a grid in \ref{section:cmae} and concludes with the evaluation metrics in \ref{section:evaluation}.

\subsection{Traffic simulation and the two sensor network scenarios}
\label{section:scenarios}

To obtain a sensor network with realistic spatial and temporal characteristics, i.e., as of vehicles moving on streets, this work utilizes a microscopic traffic simulation around the City of Erlangen with a spatial extension of $\qty{30.5}{\kilo\meter} \times \qty{30.5}{\kilo\meter}$ for probe count and traffic count data from October 7th, 2022, 11:30 to 12:30 \cite{Veihelmann.2024}. We selected this period for several reasons: First, a representative scenario should take place when a significant amount of solar irradiance is present, which is applicable around noon. Second, we did not want to choose a best-case scenario for the traffic volume. Therefore, the period is before the weekend traffic and the end of most people's working hours but after the heavy commuting hours. In addition, the selected period offers balanced traffic in the north and south directions of the street network, which we considered favorable for a "neutral" scenario without any direction-specific drifts. The resulting data consist of vehicle identifiers with their positions for time steps in a one-second resolution. 

While the whole traffic simulation area could be seen as a large sensor network and observation area, this work targets smaller sensor networks closer to existing research on local sensor networks in their spatial extension and the number of sensors. Thus, two small areas were selected from the whole traffic simulation area to provide scenarios for investigating the novel character of dynamic sensor networks in the domain of solar forecasting.

\begin{figure}[htb!]
    \centering
    \begin{subfigure}[b]{.4\textwidth}
        \centering
        \adjustbox{valign=c, height=4cm}{\includegraphics{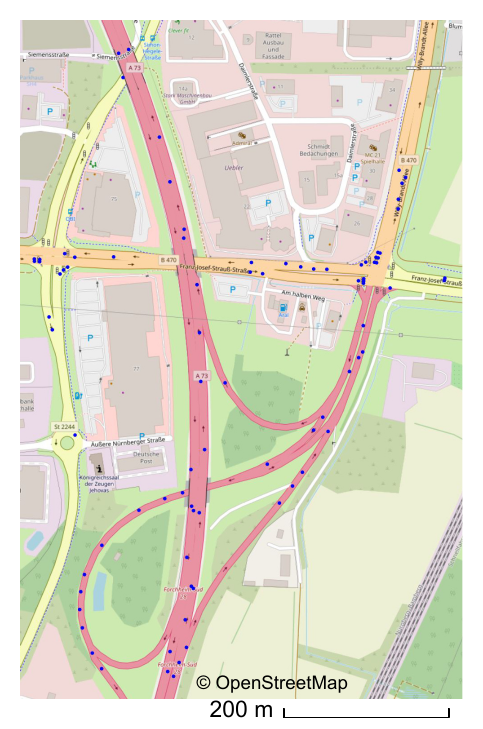}}
        \caption{Highway ramp scenario.}
        \label{fig:highwaymap}
    \end{subfigure}%
    \begin{subfigure}[b]{.4\textwidth}
        \centering
        \adjustbox{valign=c, height=4cm}{\includegraphics[width=\linewidth]{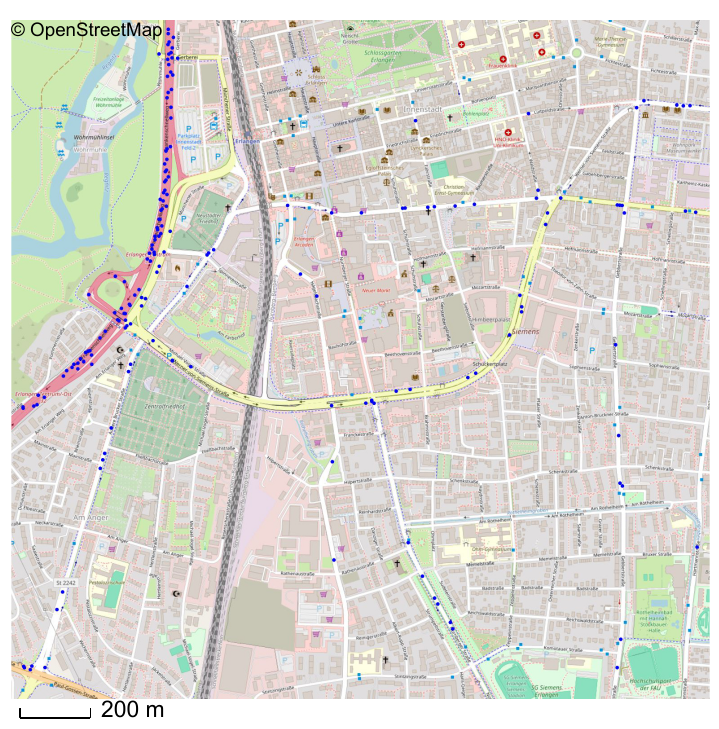}}
        \caption{Urban scenario.}
        \label{fig:urbanmap}
    \end{subfigure}
    \caption{The two observation areas with the tiny blue dots representing vehicles from a snapshot of the traffic simulation.}
    \label{fig:streetmaps}
\end{figure}

First, we identified places where the moving cars, i.e., sensors, visually appeared comparably even in their spatial distribution. Second, those areas have a small spatial extension. While the sensing performance should increase with larger observation areas, cloud dynamics in terms of cloud dissipation and formation would grow, which is not considered in this paper. Third, to investigate two sensor networks with different characteristics for improved generalizability, we created an "urban scenario" and a "highway ramp" scenario. The urban scenario includes a lot of buildings and slower-moving vehicles, while the highway ramp is at the edge of a city and mainly consists of a highway ramp with an intersection. Figure~\ref{fig:streetmaps} shows the resulting street networks. A cloud shading event lasted 300 seconds. Consequently, we extracted an excerpt from 11:50 to 11:55 from the traffic simulation. Table~\ref{tab:scenarios} provides an overview of the characteristics of the two selected sensor network scenarios. 

\begin{table}[htb!]
    \caption{Overview on the sensor networks}
    \label{tab:scenarios}
\begin{tabularx}{\textwidth}{X *{3}{C}}
	\toprule
	Scenario & Size & Median number of vehicles & Building shadows\\
	\midrule 
	Highway ramp & $\qty{0.6}{\kilo\meter} \times \qty{0.9}{\kilo\meter}$ & 83 & No \\
    Urban & $\qty{2}{\kilo\meter} \times \qty{2}{\kilo\meter}$ & 153 & Yes \\
	\bottomrule
\end{tabularx}
\end{table}

\subsection{Clear-sky index map}
\label{section:csim}

A fractal-based method by Lohmann et al. \cite{Lohmann.2017} allows the creation of a realistic image of a cloud shadow and yields a cloud-index map. For a detailed description of the underlying algorithm to create fractal-based cloud shadows, the reader may refer to \cite{Cai.2013}. As proposed by Lohmann et al. after empirically investigating fractal dimensions from real cloud images \cite{Lohmann.2017}, we set the fractal dimension to $D1 = D2 = 1.5$ in this work. Within a simulation period of 300 seconds, the cloud shadow can move up to \qty{9000}{\meter}. The maximum dimension of the larger observation tile is $\sqrt{2} \times \qty{2000}{\meter}$. Therefore, the cloud shadow pattern must be bigger than the sum of those values. Simultaneously, the power of two must be a requirement from the algorithm for generating fractal surfaces. We chose a side length of 16,384 pixels and meters to fulfill both criteria for the fractal cloud shadow field. 

We set a threshold value to distinguish shadow from non-shadow pixels to the median of the values in the array. Following \cite{Lohmann.2017}, we established a transition around the threshold of $\pm 0.15$ and linearly mapped values in the transition zone to -0.2 to 1.2.

\begin{figure}[htb!]
    \centering
    \includegraphics[width=0.5\linewidth]{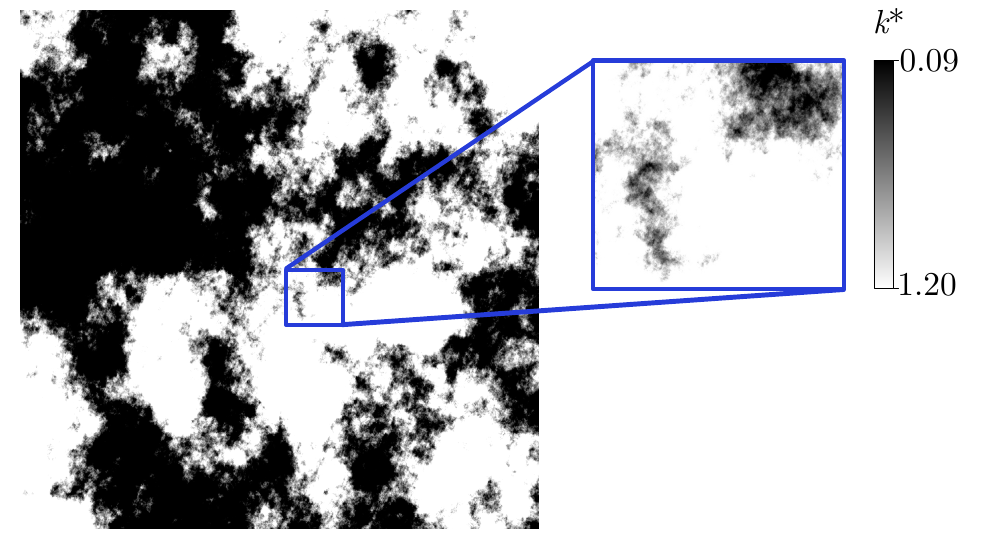}
     \caption{The cloud shadow map used for our simulation.}
    \label{fig:cloudshadow}
\end{figure}

These modeled cloud index values were then scaled down to an eight-bit representation linearly mapped to integer numbers from 0 to 255. Figure~\ref{fig:cloudshadow} shows the corresponding image. The transparency channel carries the information on shadowing intensity. While the intermediate step of converting the fractal surface values into an image may appear unnecessary from a physical point of view, they allowed us to visualize each simulation in \ref{section:transition} and validate the assignment of clear-sky index values to the vehicles' positions.
Following \cite{Lohmann.2017}, we employed the empirical relationship from Rusen et al. \cite{Rusen.2013} to transform the cloud index values $n$ into clear-sky index values $k^*$:

\begin{equation}
    k^* =
    \left\{
    \begin{aligned}
        &1.2\,, && n \le -0.2 \\
        &1 - n\,, && -0.2 < n \le 0.8 \\
        &1.1661 - 1,7814 \cdot n + 0.7250 \cdot n^2\,, && 0.8 < n \le 1.05 \\
        & 0.09\,, && n > 1.05\,.
    \end{aligned}
    \right.
\end{equation}

The clear-sky index $k^*$ describes the ratio of an irradiance component to the irradiance under clear-sky conditions \cite{Yang.2024e}. 

\subsection{Modeling the passage of cloud shadows}
\label{section:transition}

\begin{figure}[htb!]
    \centering
    \includegraphics[width=0.5\linewidth]{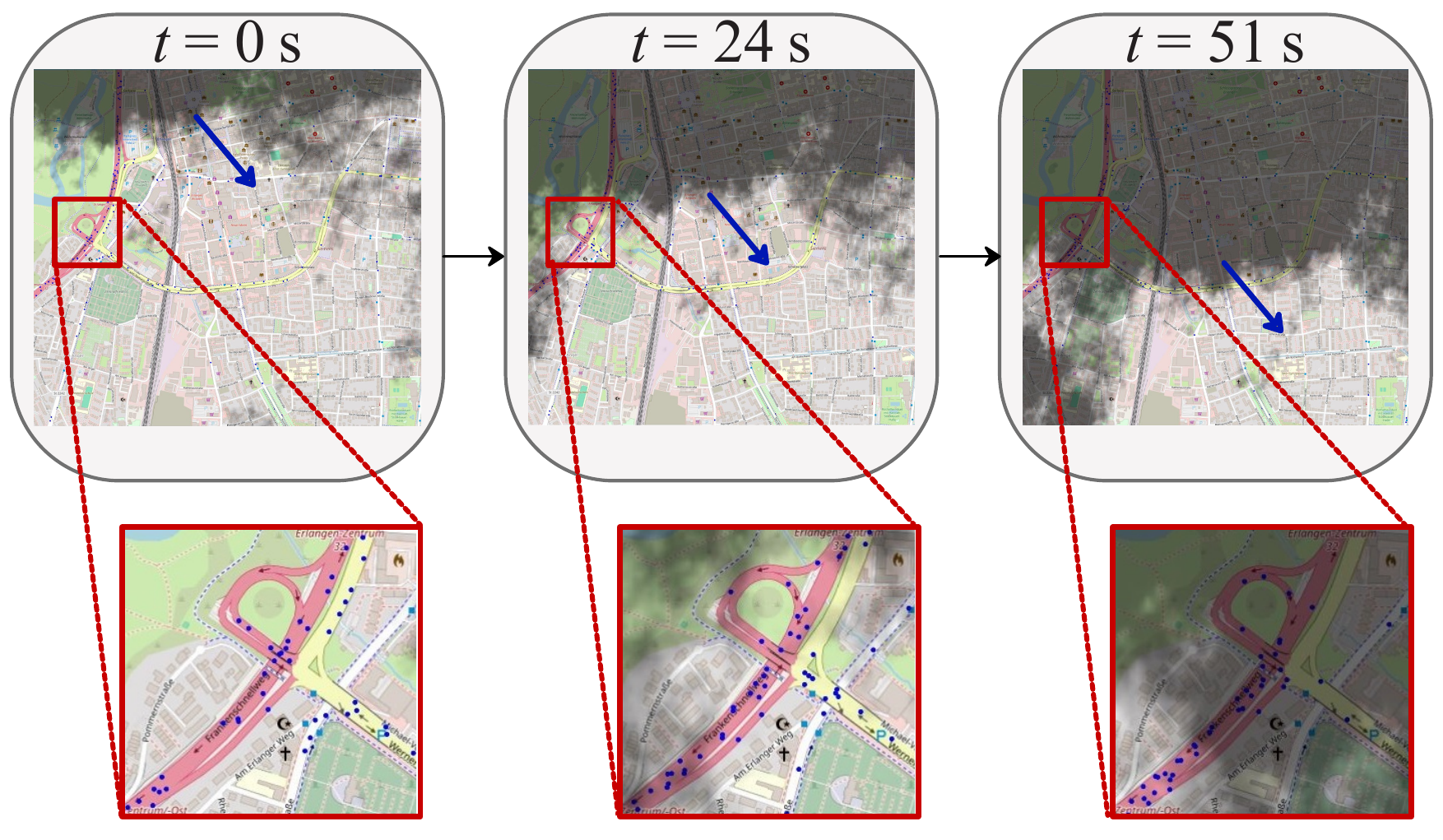}
     \caption{The cloud shadow map moving over the urban scenario's street map with the blue arrows in the upper row representing the shadow movement direction.}
    \label{fig:movement}
\end{figure}

The cloud shadow movement is modeled similarly to the methods described by Espinosa-Gavira et al. \cite{EspinosaGavira.2020}. A randomly generated cloud shadow velocity and direction are assigned to the fractal cloud model, which then moves over the traffic simulation area. The simulated velocities reach from \qty{1}{\meter\per\second} to \qty{30}{\meter\per\second}, a reasonable upper limit in line with previous works \cite{Kuhn.2018,Lappalainen.2017,Chen.2020}. The range of directions is \qty{0}{\degree} to \qty{359}{\degree}. The pixel values at vehicles' locations in the traffic simulation are captured at each sampling time. Figure~\ref{fig:movement} shows the temporal changes in both traffic simulation and cloud field over time for a cloud shadow movement simulation with a direction of 140\,° (from north) and \qty{20}{\meter\per\second}.

\subsection{Modeling shadows caused by buildings for the urban scenario}
\label{section:shadowmethods}

If applied in a dense city scenario, the proposed methods inherit an essential limitation: Many vehicles cannot sense a valid illuminance measurement that is later translated into a clear-sky index because they are obscured by building shadows. Thus, vehicles on shadowed street segments should be excluded to get a realistic view of the sensor distribution in the observation area.

\begin{figure}[htb!]
    \centering
    \includegraphics[width=0.75\linewidth]{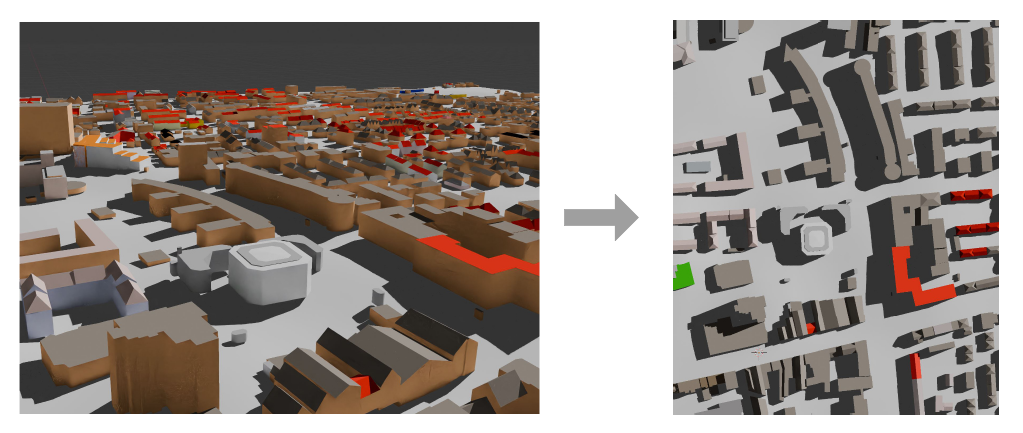}
     \caption{Excerpt of the simulated shadow map using blender with 3D data from the OpenStreetMap. The first image depicts the 3D view, the second image the 2D bird's eye view.}
    \label{fig:sunsim3D}
\end{figure}

Consequently, we utilized the 3D computer graphics software Blender. The plug-in BLOSM allowed for importing the OpenStreetMap  terrain and building files for the selected area. With the plugin Solarposition, we could model the shadows for the azimuth angle of \qty{158}{\degree} and an elevation angle of \qty{32}{\degree}, matching the datetime of the traffic simulation. Figure~\ref{fig:sunsim3D} depicts a 3D view of an excerpt of the buildings and shadows, as well as a part of the exported two-dimensional shadow map.

Subsequently, we exported a bird's eye view image file from Blender and mapped the pixels to georeferenced coordinates. For the cloud shadow transition simulations in the urban scenario, only vehicles at non-shadowed pixel locations were considered as sensors at each sampling time in the simulation. After excluding those vehicles, a median of 122 vehicles from the median number of 153 remained in the urban scenario.

\subsection{Applied CMAE method}
\label{section:cmae}

To derive cloud shadow velocities and directions from time series of measured irradiance, we employed the CMAE method, which requires time series for a grid of sensors. Espinosa-Gavira et al. \cite{EspinosaGavira.2020} showed how their CMAE method could more accurately sense cloud shadow velocities on gridded small-scale sensor networks compared to the more established linear cloud edge (LCE) \cite{Bosch.2013} and most-correlated pair (MCP) \cite{Bosch.2013} method. 

\begin{figure}[htb!]
    \centering
    \includegraphics[width=0.75\linewidth]{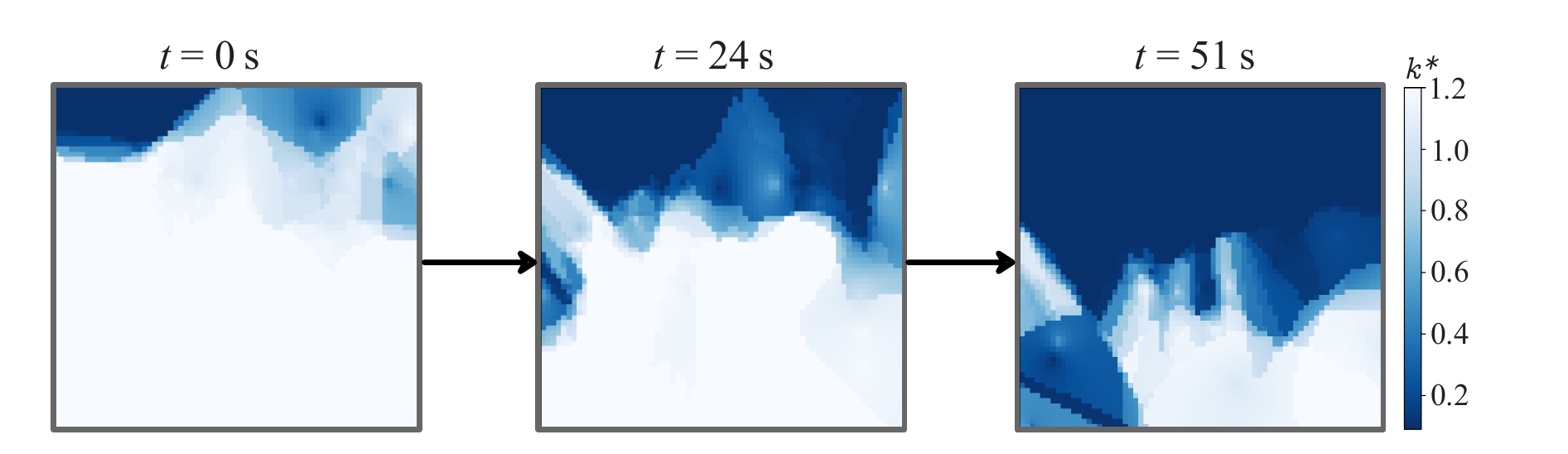}
     \caption{The measurements from the sensors in Figure~\ref{fig:movement} after applying inverse-distance weigthing to the three-nearest neighbors with a grid point distance $D$ of \qty{30}{\meter}.}
    \label{fig:interpol}
\end{figure}

With the sensors comprising the network moving and not being evenly distributed over the observation area, the CMAE method cannot directly be applied to the presented sensor network. First, the measurements needed a translation into a grid, done here with an inverse-distance weighted interpolation (IDW). For each sample, the three nearest vehicles to each grid point are weighted according to their inverse distance from the grid point and the resulting interpolated value is assigned to the grid point. Figure~\ref{fig:interpol} shows the resulting interpolated clear-sky indices for a grid with a minimum grid point distance of \qty{30}{\meter}. Subsequently, the CMAE method can be applied without any modifications.

Similarly to calculating a cross-correlation of two two-dimensional arrays, the CMAE method compares two sensor grids' measurements for different two-dimensional displacements. In contrast with cross-correlation, no convolution is performed, but instead, the values are subtracted to yield absolute errors. Those are normalized with the number of "overlapping" pixels/sensors. The formula for calculating the MAE for a given displacement vector for two snapshots $S$ from the times $t$ and $t+T_s$ as introduced by \cite{EspinosaGavira.2020} is 

\begin{equation}
\label{eq:maeesp}
\mathrm{MAE} = \frac{1}{N} \sum_{k=1}^N | S'_{t}(k) - S'_{t+T_s}(k) |\,.
\end{equation}

$S'$ represents the respective snapshot parts considering the displacement for which the MAE is calculated with $N$ being the size of $S'$.

Repeating and accumulating this calculation over the whole 300-second observation period results in the cumulative mean absolute error. Calculating the CMAE for each possible displacement, selecting the three displacements with the lowest CMAEs, and inverse-distance weighting them yields the velocity and direction estimate \cite{EspinosaGavira.2020}. To exclude physically implausible estimates, we only consider displacements corresponding to velocities $\leq$ \qty{40}{\meter\per\second}.

\subsection{Evaluation metrics}
\label{section:evaluation}

The root mean square error (RMSE) is the evaluation metric for aggregating the CMAE method's direction and velocity estimates of the 100 cloud shadow movement simulations. RMSE is chosen as the evaluation metric since the errors in velocity and direction estimates are normally distributed, which is the RMSE's underlying assumption \cite{Chai.2014}. The RMSE is calculated as follows \cite{Chai.2014}:

\begin{equation}
\text{RMSE} = \sqrt{\frac{1}{n} \sum_{i=1}^n {\epsilon_{i}}^{2}}\,,
\end{equation}

where there are $n$ samples of model errors $\epsilon$.

However, even a single significant outlier strongly impacts the RMSE metric with our sample size. Thus, we do not rely on the RMSE metric alone for the analyses but visualize the errors with scatter plots to have more informed insights and understand if RMSE changes are driven by the population of simulations or by isolated but influential events.

\section{Results}

\subsection{Considerations on spatio-temporal dependencies}

With a minimum grid point distance of \qty{10}{\meter}, a velocity difference in \qty{1}{\meter\per\second} will require a time step of \qty{10}{\second} to result in a shift of one pixel in the interpolated clear-sky index map. A time step denotes the time difference between subsequent snapshots of measurements from the sensor network as input for the CMAE method. Because the CMAE methods returns the inverse-distance weighted aggregation of three displacements, the method may in practice still output slightly higher-resolved estimates. To find a limit where the estimates would start not to be coarsely binned, we used time step combinations where the minimum grid point distance $D_\text{min}$ in meters was ten times as high as the time step in seconds as an upper limit; i.e., in theory, velocity differences of \qty{10}{\meter\per\second} could be recognized. 

Since the highway ramp scenario's dimensions are $\qty{620}{\meter} \times \qty{930}{\meter}$, for maximum velocities to be estimated, the maximum time step is $\qty{620}{\meter} \div \qty{30}{\meter\per\second} = \qty{20.7}{\second}$ (shown for up to \qty{15}{\second}). The dimensions of the city centre scenario are $\qty{2}{\kilo\meter} \times \qty{2}{\kilo\meter}$. Subsequently, the upper limit for the time steps is $\qty{2000}{\meter} \div \qty{30}{\meter\per\second} = \qty{66.7}{\second}$. Consequently, we selected time steps from \qty{10}{\second} to \qty{60}{\second} (shown for up to \qty{30}{\second}) in \qty{10}{\second} steps. With an approximate factor three difference in the shorter side lengths of the two scenarios, we also increase the minimum grid point distance from \qty{10}{\meter} to \qty{30}{\meter}. For the same reason and \qty{2}{\second} being a divisor of all time steps in the urban scenario, we set the shortest sampling period to \qty{2}{\second} instead of \qty{1}{\second}.

Since the CMAE method relies on analyzing variability in irradiance, only simulations with some minimal irradiance variability due to cloud shadow movement shall be considered for further analysis. We divided the two sensing areas into nine equally sized rectangles. Only simulations where irradiance changes in the central rectangle had been detected for more than \qty{60}{\second} (i.e., at least one sensor in the area has detected a change in irradiance) were considered for further analysis. After this step, 89 of the 100 simulations remained for the highway ramp and 97 for the urban scenario.

\begin{figure}[htb!]
    \centering
    \begin{subfigure}{.5\textwidth}
        \centering
        \includegraphics[width=\linewidth]{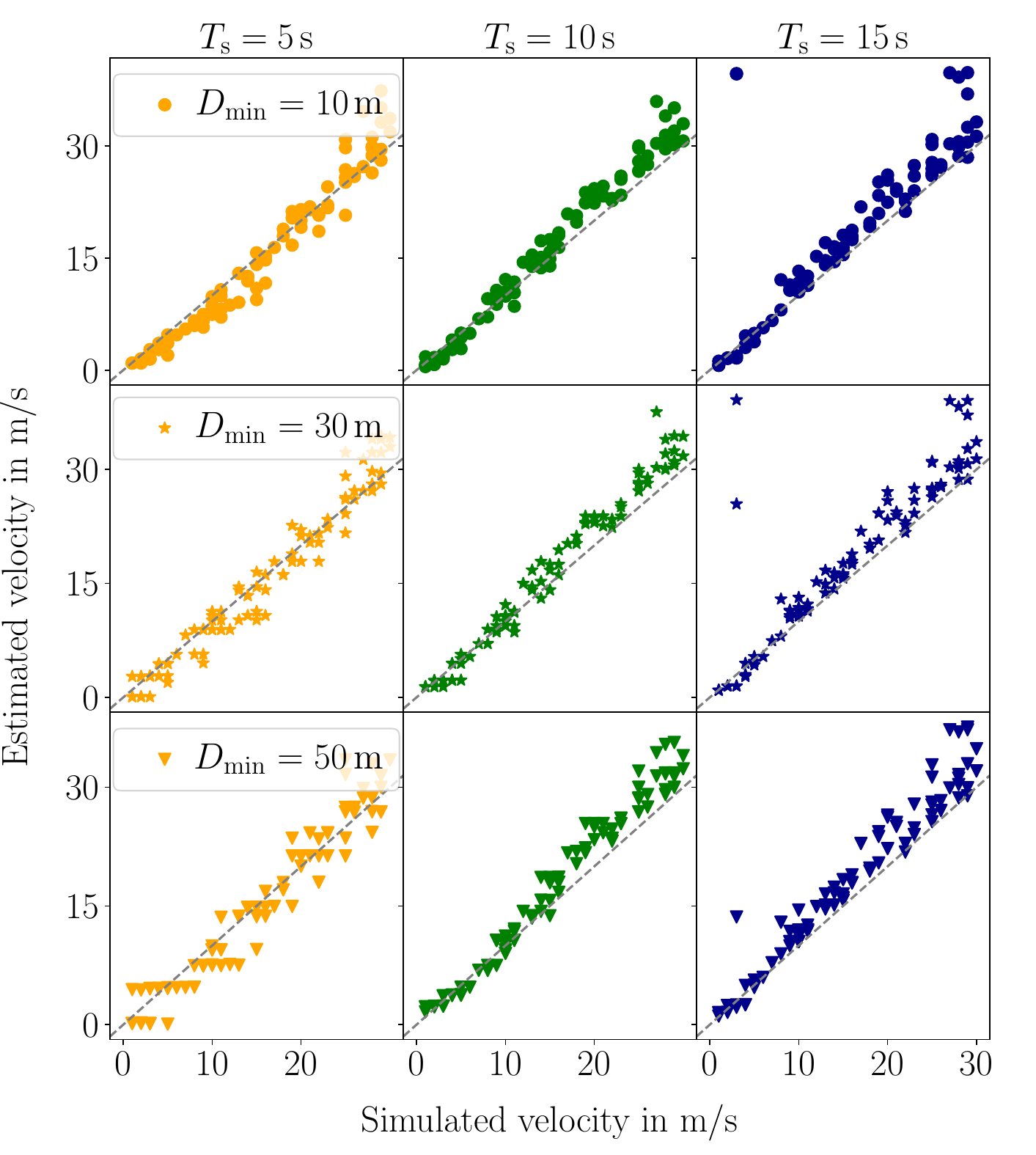}
        \caption{Highway ramp scenario.}
        \label{fig:highwayvelocity}
    \end{subfigure}%
    \begin{subfigure}{.5\textwidth}
        \centering
        \includegraphics[width=\linewidth]{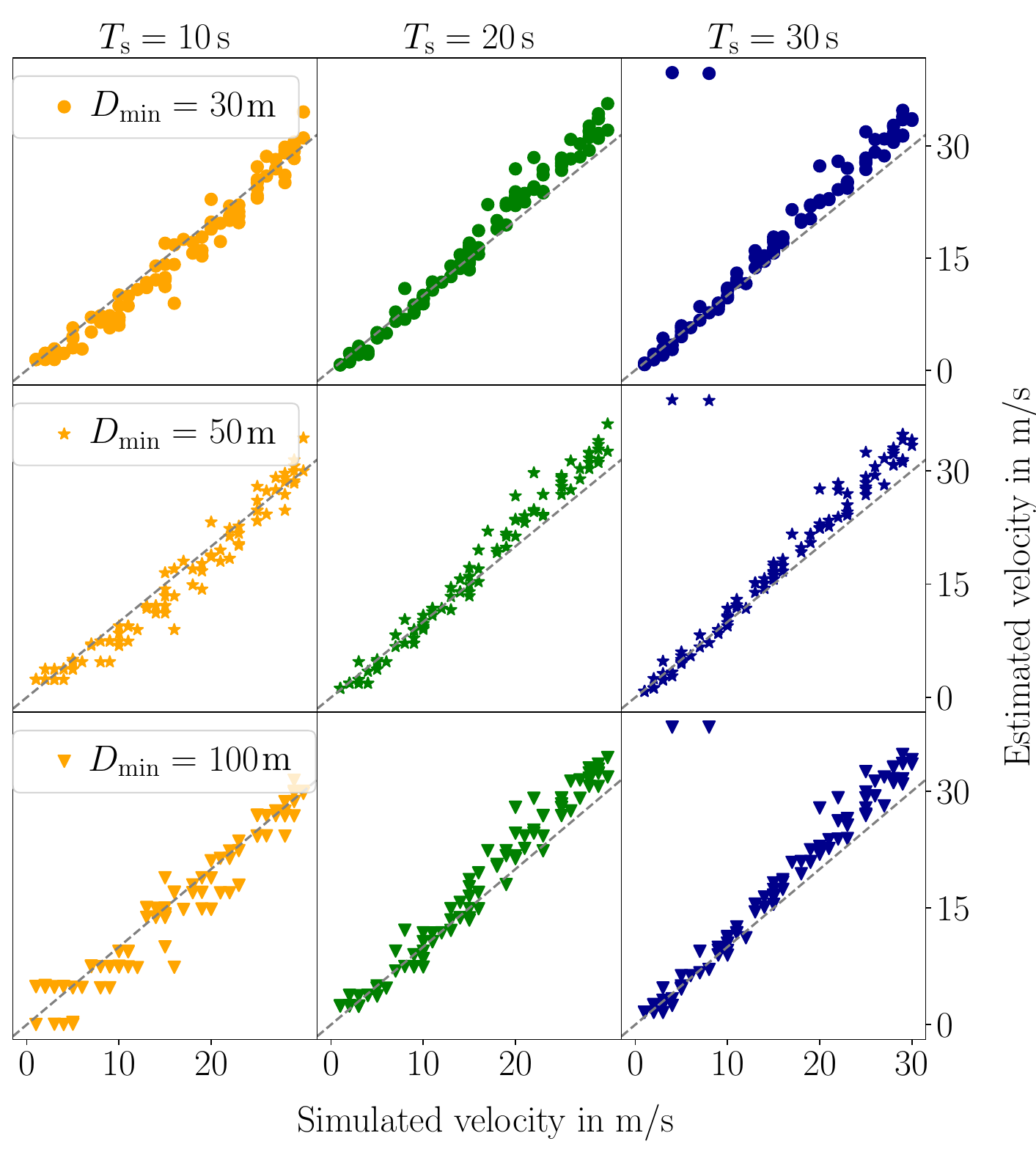}
        \caption{Urban scenario.}
        \label{fig:urbanvelocity}
    \end{subfigure}
    \caption{Simulated and estimated velocities for both scenarios. The sampling period for the highway ramp scenario is \qty{1}{\second}, for the urban scenario it is \qty{2}{\second}.}
    \label{fig:speedsimulation}
\end{figure}

For the highway ramp scenario, the results show that there is significant binning for estimated velocities with time steps below \qty{15}{\second} for minimum grid point distances of \qty{30}{\meter} or \qty{50}{\meter}. A better resolution for velocity prediction is achieved with time steps of \qty{15}{\second} or \qty{20}{\second}. However, time steps of \qty{5}{\second} or \qty{10}{\second} yield the most stable results, suggesting that a grid point distance of \qty{10}{\meter} may offer a good compromise of low time steps and high resolution. Comparing the results for a grid point distance of \qty{10}{\meter}, the results for a time step of \qty{10}{\second} look more stable. 

While the results for higher time steps appear less stable, the estimates for lower velocities precisely match the simulated velocities, indicating that the accuracy of velocity estimates steadily increases for longer time steps. This trend is present for all grid point distances. 

The urban scenario shows a similar picture: The lowest time step of \qty{10}{\second} in combination with higher $D_\text{min}$ of \qty{50}{\meter} or \qty{100}{\meter} leads to visible binning in the simulated and estimated velocity plots in Figure~\ref{fig:urbanvelocity}. Inspecting the shape of the scatter plot for a $D_\text{min}$ of at least \qty{10}{\meter} and a time steps exceeding \qty{5}{\second}, we find the same pattern as \cite{EspinosaGavira.2020} when they applied Kriging to a non-gridded sensor network: While the predictions seem unbiased for lower velocities, the CMAE method tends to overestimate higher velocities exceeding \qty{15}{\meter\per\second}. The same phenomenon is present in the urban scenario for time steps of at least \qty{20}{\second} for all $D_\text{min}$.

\begin{figure}[htb!]
    \centering
    \begin{subfigure}{.5\textwidth}
        \centering
        \includegraphics[width=\linewidth]{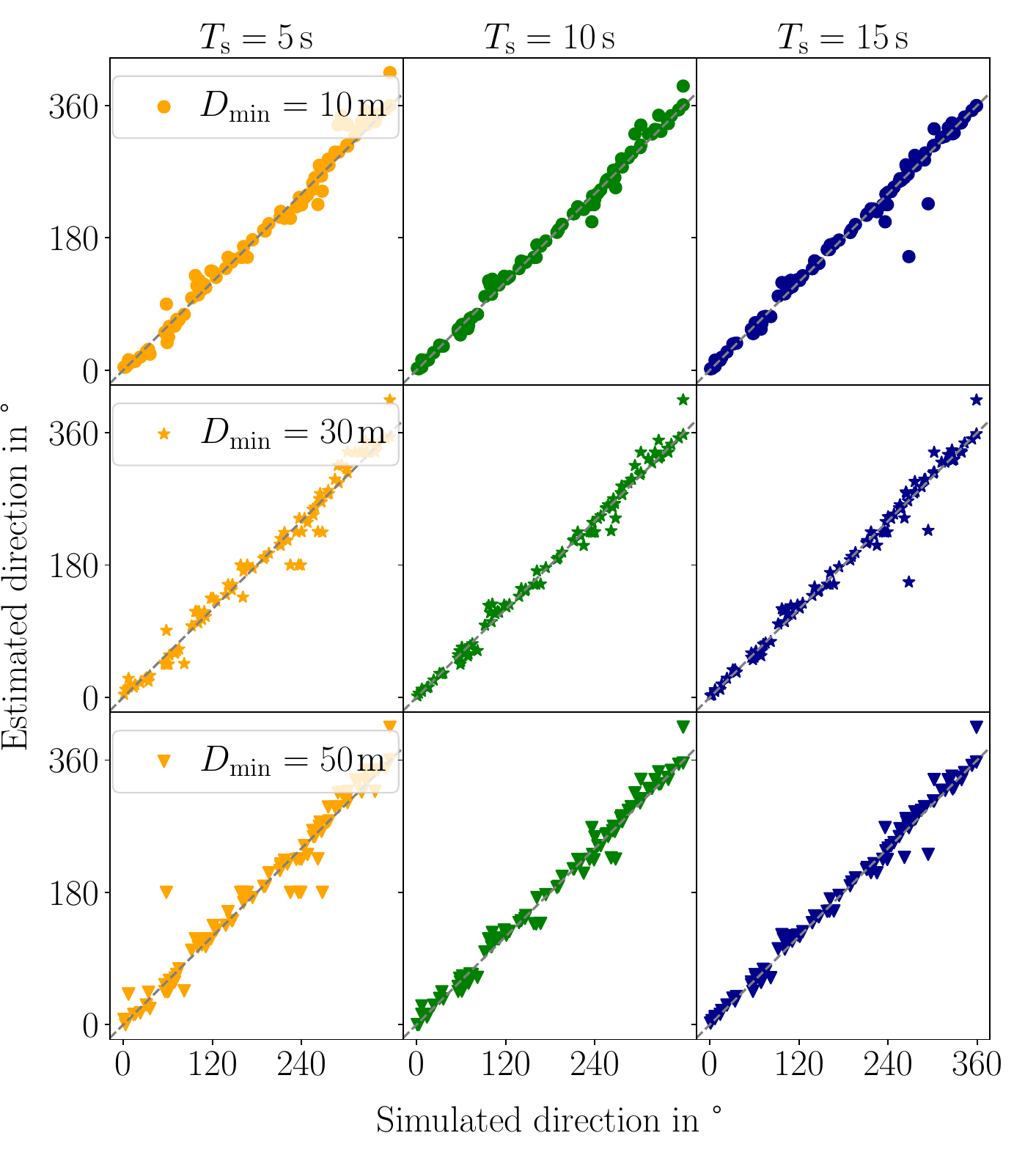}
        \caption{Highway ramp scenario.}
        \label{fig:highwaydireciton}
    \end{subfigure}%
    \begin{subfigure}{.5\textwidth}
        \centering
        \includegraphics[width=\linewidth]{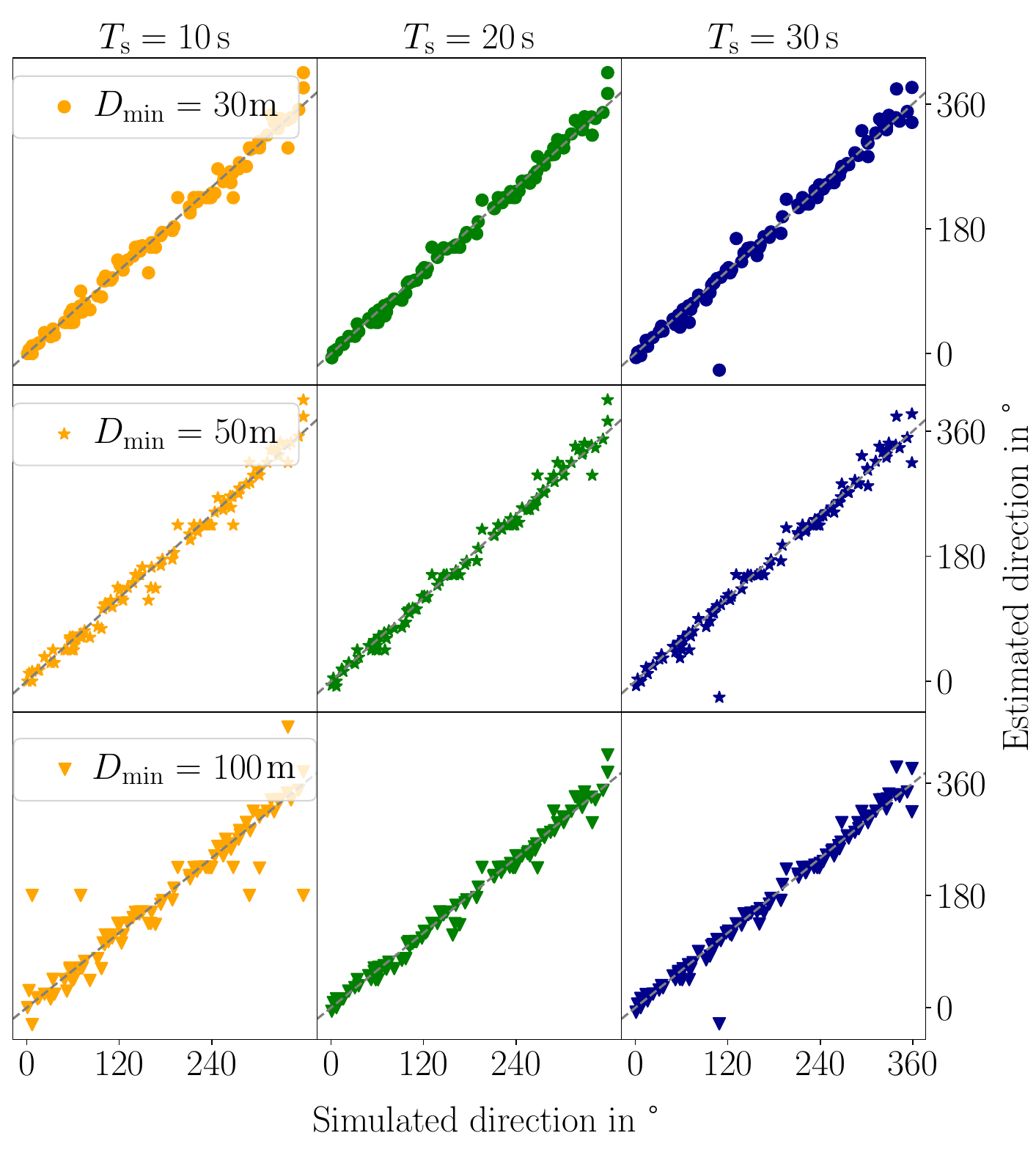}
        \caption{Urban scenario.}
        \label{fig:urbandirection}
    \end{subfigure}
    \caption{Simulated and estimated directions for both scenarios. The sampling period for the highway ramp scenario is \qty{1}{\second}, for the urban scenario it is \qty{2}{\second}.}
    \label{fig:direction}
\end{figure}

The pattern for direction estimates in Figure~\ref{fig:direction} appears quite similar to the velocity estimate pattern with two differences: First, the limited resolution for velocity estimates with a time step of \qty{10}{\second} and grid point distances of at least \qty{30}{\meter} is not apparent for direction estimates. Second, the outliers that appear for time steps of \qty{15}{\second} look less dramatic - while they are up to \qty{46}{\degree} off, they still seem to go somewhat into the right direction, which cannot be stated for the bounded velocity estimates.

\subsection{Reducing the number of vehicles}

So far, we have assumed that all vehicles modeled in the traffic simulation are equipped with communication capabilities, light sensors, and that their data would be available. This subsection shows how the RMSEs increase if less vehicles' measurements are available. The shares of vehicles, i.e., vehicle identifiers, from the traffic simulation considered for cloud shadow sensing vary from 10\,\% to 90\,\% in 10\,\% increments. In the following, we refer to the share of vehicles used for sensing, i.e., assumed to have both communication and sensing capabilities and are willing to share their data, as the penetration rate. We employed the parameters in Table~\ref{tab:hypercs} for the following analyses. 

\begin{table}[htb!]
    \caption{Parameters for analysis on penetration rates}
    \label{tab:hypercs}
\begin{tabularx}{\textwidth}{X *{4}{C}}
	\toprule
	Scenario & $D_\text{min}$ in m & Time step in s & $N_\text{Neighbors}$ for IDW \\
	\midrule 
	Highway ramp & 10 & 10 & 3\\
    Urban & 30 & 20 & 3\\
	\bottomrule
\end{tabularx}
\end{table}

\begin{figure}[htb!]
    \centering
    \begin{subfigure}{.42\textwidth}
        \centering
        \includegraphics[width=\linewidth]{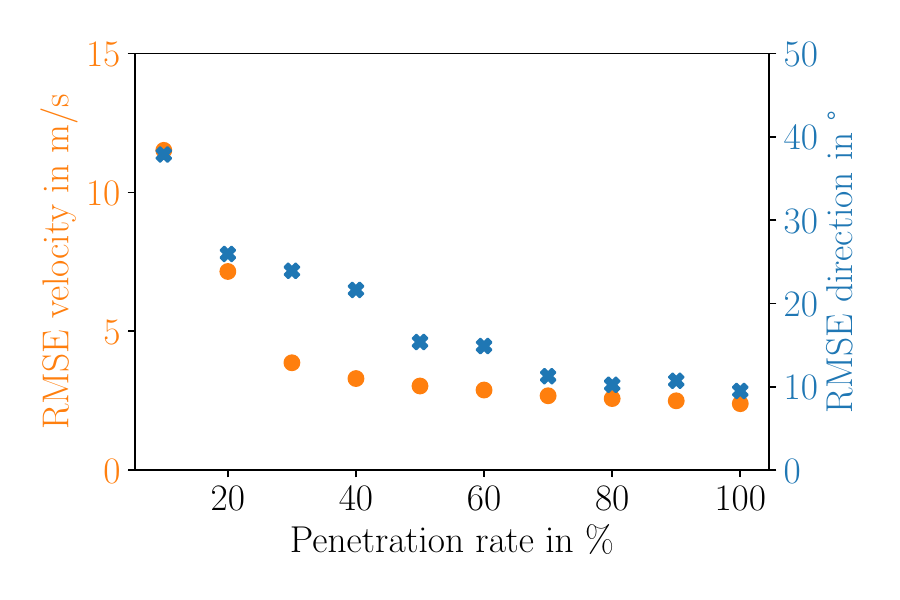}
        \caption{Highway ramp scenario.}
        \label{fig:focsbase}
    \end{subfigure}%
    \begin{subfigure}{.42\textwidth}
        \centering
        \includegraphics[width=\linewidth]{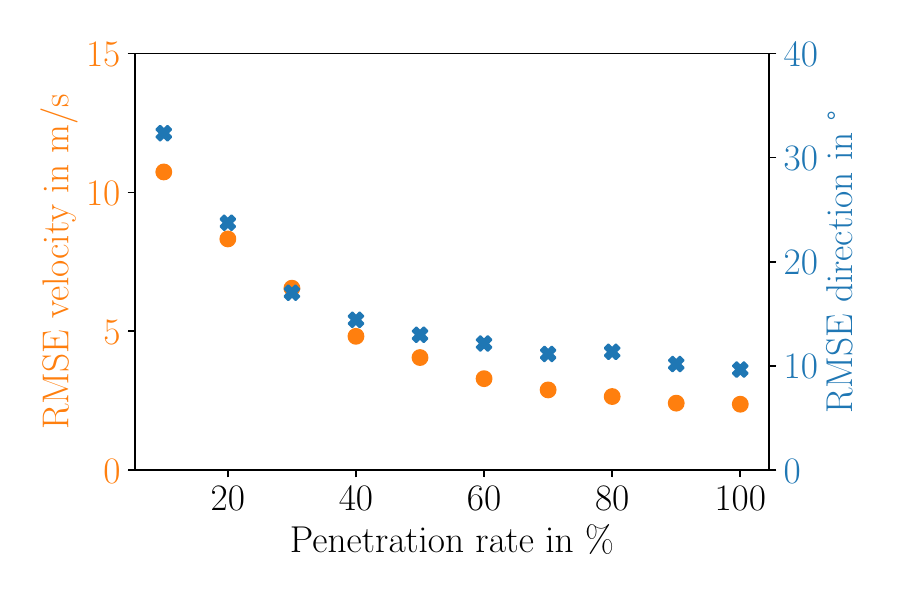}
        \caption{Urban scenario.}
        \label{fig:ercsbase}
    \end{subfigure}
    \caption{RMSEs for velocity and direction estimate for both scenarios over varying penetration rates.}
    \label{fig:csbasermse}
\end{figure}

Figure~\ref{fig:csbasermse} shows for both scenarios that a penetration rate of \qty{10}{\percent} leads to significant errors for the velocity estimates. However, a solid correlation between estimated and simulated velocities starts with a penetration rate of \qty{30}{\percent} and an RMSE of \qty{3.9}{\meter\per\second} for the highway ramp scenario and with a penetration rate of \qty{40}{\percent} and an RMSE of \qty{4.8}{\meter\per\second} for the urban scenario. While the changes from penetration rates from \qty{50}{\percent} to \qty{90}{\percent} appear marginal for the highway ramp scenario with RMSE improvements from \qty{3.0}{\meter\per\second} to \qty{2.5}{\meter\per\second}, the urban scenario's velocity estimates gradually improve until a penetration rate of \qty{90}{\percent} of traffic simulation vehicles included in the analysis with an RMSE of \qty{2.4}{\meter\per\second}. However, the urban scenario's estimates may be handicapped by some vehicles being excluded due to building shadows and the generally smaller number of additions for the CMAE method due to the longer sampling period of \qty{2}{\second}. Thus, the usually larger observation area and higher number of vehicles in the analysis seem to overcompensate those disadvantages.

The RMSEs for direction estimates tell a slightly different story: here, the urban scenario performs slightly better than the highway ramp scenario. While there are no severe outliers from a penetration rate of \qty{40}{\percent} and above for the urban scenario, outliers in the highway ramp scenario disappear only with a \qty{50}{\percent} penetration rate with an RMSE of \qty{15.4}{\degree}. The agreement of simulated and estimated directions appears very good with at least \qty{50}{\percent} of sensing vehicles in the urban and at least \qty{60}{\percent} in the highway ramp scenario. For both scenarios, the performance further increases until \qty{100}{\percent} of sensing vehicles with RMSE of \qty{9.5}{\degree} (highway ramp) and \qty{9.7}{\degree} (urban), respectively.

While the plots in Figure~\ref{fig:csbasermse} indicate that the estimation of cloud dynamics suffers from not having all cars on the streets available for illuminance or irradiance sensing, having \qty{40}{\percent} of vehicles available for sensing offers a very good relation between true and estimated velocity and direction estimates, which is also supported by the RMSE values. The corresponding scatter plots and tables with RMSE values are provided in the supporting information.

\subsection{Varying the time of day for the urban scenario}

Since the previous results apply all to the same time of - an October - day around noon when shadows have their lowest extents, we also investigated how the results of the urban scenario would change for two alternative times: One additional scenario targets 2:50 p.m. when the elevation angle is \qty{30.2}{\degree} - hence, pretty similar to the reference one with \qty{32}{\degree} - but the traffic volume is significantly increased to a median of 230 (without building shadows: 203) active vehicles. The second additional scenario takes place at 4:50 p.m. with a significantly lower elevation angle of \qty{16}{\degree} and a higher median traffic volume of 226 (without building shadows: 163) active vehicles. These numbers result from additional traffic simulations we produce with the procedure described in \cite{Veihelmann.2024} with the respective traffic count numbers for the changed times of day.

\begin{figure}[htb!]
    \centering
    \includegraphics[width=.99\linewidth]{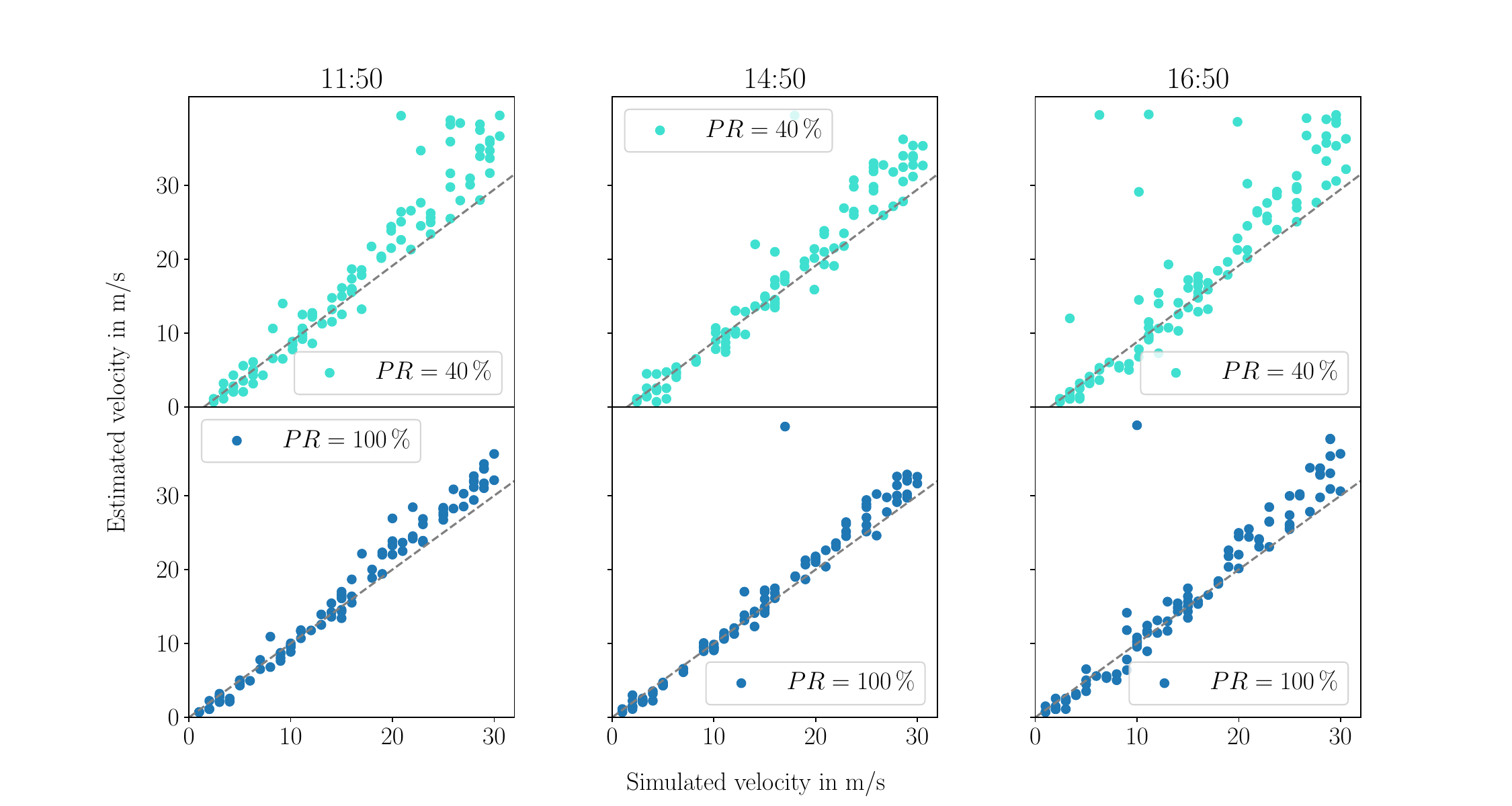}
    \caption{Simulated and estimated velocities with penetration rates ($PR$) of \qty{40}{\percent} and \qty{100}{\percent} for three different times of day.}
    \label{fig:todcomp}
\end{figure}

Figure~\ref{fig:todcomp} shows scatter plots of simulated and estimated velocities for the different times of day for the penetration rates of \qty{40}{\percent} and \qty{100}{\percent}. For both penetration rates, the simulation at 2:50 p.m. generally shows the closest agreement of both values. However, over all penetration rates (also the ones not shown), one significant outlier has a predicted velocity of almost \qty{40}{\meter\per\second}, hitting the limit of the maximum velocity filter.

The simulations at 4:50 p.m. produced many outliers for the lower penetration rates, indicating that the robustness of the CMAE method is significantly challenged by the combination of many cars in the city center being obscured by shadows and an overall lower penetration rate. The higher number of vehicles at this time compared to the reference scenario indicates that the distribution of vehicles is more important than their total amount. Recalling the urban scenario in Figure~\ref{fig:urbanmap} illustrates that most vehicles are on the highway section on the top left of the map. Excluding the shaded cars has its main effect in the city center, resulting in the overall number of vehicles being higher than at 11:50 a.m. However, most cars are active on the line topology of the highway, not providing as much spatial entropy as the sensor network at 11:50 a.m.

For the estimated directions, there are fewer errors for simulated times. Only the 4:50 p.m. scenario shows two outliers. Tables with RMSE values for velocity and direction estimation, as well as examplary scatter plots for the direction estimates are provided in the supporting information. These show that the direction estimates for 2:50 p.m. are improved compared to the reference time, which is also reflected in the RMSE values since there is no outlier for direction estimates. For the 4:50 p.m. scenario, however, there is an outlier, causing the RMSE to never fall below \qty{18}{\degree}. With the lower penetration rate of \qty{40}{\percent}, there are two outliers, increasing the RMSE to \qty{26.1}{\degree}. The RMSE tables reveal that for the 11:50 a.m. and 2:50 p.m. simulations there is a continuous improvement in RMSE values for all penetration rates from \qty{40}{\percent} to \qty{100}{\percent}. For the 4:50 p.m. scenario, however, there is a significant improvement when increasing the penetration rate from \qty{40}{\percent} to \qty{50}{\percent} before the errors diminish similarly to the other times of day.

\section{Discussion}

\subsection{Overall findings and performance}

The results provide convincing evidence that - under the assumption of perfect sensors - the CMAE method, in conjunction with the proposed methods, can provide reasonably accurate cloud shadow velocities and directions. In a future scenario with all cars being connected and equipped with very accurate irradiance sensors, the distribution in an urban scenario of the German City of Erlangen at noon could enable to sense cloud shadow velocities with an RMSE of \qty{2.4}{\meter\per\second} and directions with an RMSE of \qty{9.7}{\degree}. While analyses with a penetration rate of \qty{100}{\percent} may be far from realistic, a penetration rate of \qty{40}{\percent} would still suffice for an RMSE of \qty{4.8}{\meter\per\second} and \qty{14.4}{\degree}. 

The second, smaller scenario with fewer vehicles still yields RMSE values of \qty{2.4}{\meter\per\second} and \qty{9.8}{\degree} for full penetration and \qty{3.3}{\meter\per\second} and \qty{21.6}{\degree} for \qty{40}{\percent} penetration rate. A penetration rate of \qty{40}{\percent} is realistic as in Germany, for example, the Volkswagen AG brands VW, Audi, Porsche, Skoda, and Seat had a cumulated market share of \qty{38.7}{\percent} in 2023 with a similar share for new registrations \cite{VDIK.2024}. 

While the RMSE values steadily diminish with increased penetration rates, the biggest improvements happen from enhancing penetration rates from \qty{10}{\percent} to \qty{40}{\percent}. Depending on the sizes of so-thought observation areas, the number of CMAE results with larger errors strongly depends on the selected time step, which must not be too long. The errors rise before the upper theoretical limit is reached, which, however, can be explained by vehicles not (necessarily) present at the very fringes of the observation areas. The minimum grid-point distance does not influence the sensing errors significantly. Inspecting the plots suggests that larger grid point distances lead to significant binning of velocity estimates and, hence, reduced precision.

\subsection{Comparison to static grid(s)}

In this subsection, we refer to the results for all cars included in sensing activities. Comparing the results to the results in \cite{EspinosaGavira.2020} may be slightly biased due to the different logics of defining a valid cloud movement event: Their criterion was that at least one sensor in the network needed to measure irradiance changes for 150 seconds. Our simulations, in contrast, required 60 seconds of irradiance variation detected by at least one sensor, but in the central square of the observation area. This change was necessary since the sensors are moving and the grids are more extensive, resulting in broader grid point distances and smaller likelihoods of a second sensor measuring variability if this applies to one sensor.

Still, even for the smaller network, 89 of 100 simulations are used for subsequent analysis in our case. \cite{EspinosaGavira.2020} yielded their best results for a grid with 100 sensors aligned in an even grid over an area of $\qty{30}{\meter} \times \qty{30}{\meter}$ with an RMSE of \qty{1.4}{\meter\per\second}. The urban scenario in our work comes close with \qty{2.4}{\meter\per\second} with a median of 122 moving vehicles. However, while the distribution of errors shown in \cite{EspinosaGavira.2020}'s base scenario does not depend on the simulated velocity, the results in our paper show a trend increasing toward higher velocities. This pattern resembles an analysis in \cite{EspinosaGavira.2020} where they applied kriging to transfer arbitrarily distributed sensors to a grid. One could argue that there may be a slight tendency for this pattern to be amplified in this work since lower velocities may be slightly underestimated, which cannot be observed in \cite{EspinosaGavira.2020}.

While \cite{EspinosaGavira.2020} did not provide RMSEs for direction estimates, the overall patterns compare well. The plotted relations of simulated and estimated directions resemble the ones in this work, with the only difference being that slightly more outliers appear in their work. First, our criterion for valid simulations may be slightly more restrictive. Second, while we closely followed \cite{Lohmann.2017} for the generation of the clear-sky index map, \cite{EspinosaGavira.2020} used a slightly different mapping from fractal surface to clear-sky indices with a lower difference between the highest and lowest clear-sky index values. Third, the scenarios' spatial extension significantly exceeds the grid layouts tested by \cite{EspinosaGavira.2020}. With very similar fractal dimensions and, consequently, cloud shadow shapes as in \cite{EspinosaGavira.2020}, a more extensive observation area includes more variation and a bigger picture of differences. Thus, our results may benefit from the larger area, which is always preferable if there is still one dominant direction and velocity as a ground truth to be detected.

\subsection{Scalability and improvements of the approach}

What should be considered when judging the results presented in this paper is the scalability of the presented approach. While conventional local sensor networks are limited by the area (see Table~\ref{tab:lsn}) where the sensors are deployed - and their number - the approach presented here can be extended to all continental areas with street networks. Of course, the resolution is limited by the number of vehicles whose data can be used for sensing, but increasing the expansion of the network may also reduce errors in velocity and direction estimates even without having a higher resolution of vehicles. Since clouds may gradually vary their velocity and direction, a bigger observation area may provide a priori knowledge before the cloud field enters the "actual" observation area of interest.

Yet the scalability potential does not only apply to the spatial dimension of the network: establishing a virtual sensor grid also allows the inclusion of various types of additional dynamic and, of course, static sensors that can be interpolated and used in addition to the automotive light sensors. For example, private weather stations with communications capabilities, mobile radio base stations with irradiance sensors, and obviously "real" weather stations could be integrated into such a framework.

Furthermore, applying the CMAE method is just a starting point for considering moving vehicles for approaches to cloud motion estimation. Although the results - achieved with the assumption of perfect sensors - appear convincing, it is not very plausible that the base CMAE method is the best possible approach to realize this. While the interpolation to a virtual sensor grid is a pragmatic and solid first step for data preparation, there may be more sophisticated interpolation approaches that may lead to better results. In addition, the CMAE method considers all grid points equally. In our street network, line topology-based scenarios show that some grid points are near streets where vehicles can be present, and others are very far from any roads. Thus, we see some weighting or pre-selection of grid points to include for the CMAE method as first steps towards improvement and, ultimately - with careful creation of scenarios and sampling - machine learning-based approaches as promising further developments. To support the development of novel approaches for cloud motion estimation for dynamic sensor networks, we provide our simulation data generated in subsection~\ref{section:shadowmethods} to \ref{section:cmae} in a public repository \cite{Veihelmann.2024b}.

\subsection{Limitations}

Despite carefully applying established methods from the solar forecasting literature, some limitations apply to this work. That the sensing accuracy of automotive light sensors for irradiance sensing remains unknown is an underlying assumption in this paper rather than a classical limitation. In addition, the aim to establish a theoretical upper limit of the potential of the proposed approaches should not be restricted. However, this assumption does not only imply that we did not model any sensor errors but also that there is no variability in the sensor accuracies. Those unknown effects should be subject to future investigations focusing on the distribution of such errors. Still, with the inaccuracies due to moving vehicles and interpolation to grid points, there is already some evidence of robustness in the proposed approaches. Thus, this work provides a valid case that it is worth pursuing such investigations and potentially developing post-processing methods correcting for sensing inaccuracies.

Another limitation lies in the static nature of the fractal cloud shadow field utilized in this work. While clouds, in reality, are constantly subject to formation and dissipation processes, this work assumes a static cloud field over the simulation period of 5 minutes and the observation areas. Although practically, the constraints are significantly smaller since each cloud shadow part is only assumed to be fixed during its movement over the observation area (which is typically way shorter than 5 minutes), they are further relaxed by the fact that the edges become "softer" anyways due to the interpolation to grid points. In addition, the observation areas in this paper are significantly bigger than the $\qty{100}{\meter} \times \qty{100}{\meter}$ in \cite{EspinosaGavira.2020} but still not measuring many square kilometers, we argue that these effects do not significantly skew the results. Replications with irradiance maps derived from LES might add valuable insights.

Not least, our building shadow modeling is not entirely accurate due to our areal two-dimensional simulation. We excluded vehicles at the obscured pixels representing the ground on the shadow image. Current automotive light sensors are located at the top of a vehicle's windshield, implying not the shadow at ground level but at a height of approximately \qty{1.5}{\meter} should be considered. However, we see this limitation not as too problematic for this study since, with vehicles modeled as a single coordinate and pixel sizes of \qty{1}{\square\meter}, and the background of modeling more than 100 vehicles on an area of up to $\qty{2}{\square\kilo\meter} \times \qty{2}{\square\kilo\meter}$, the overall simulation architecture does not rely on sub-meter accuracy in 3D space on this tiny scale. Furthermore, excluding vehicles based on ground shadings is a more restrictive, not a more relaxed criterion and may, in the worst case, harm and not benefit the results.

\section{Conclusion}

This work has provided a framework for how accurate irradiance values from the positions of moving connected vehicles could be leveraged for cloud motion estimation. Converting proposals to interpolate randomly distributed measurements to an even grid with inverse-distance weighting allows the implementation of the CMAE method to a dynamic sensor network. Assuming complete penetration, the results of 89 and 97 simulations from two simulative observation areas come close to the comparable simulations for static networks. This result shows the generalizability of the CMAE method for dynamic sensor networks as well as the basic suitability of irradiance measurements from vehicles for cloud motion estimation tasks. Lowering the penetration rates to \qty{40}{\percent} increases the errors in cloud motion estimates. Further lowering the penetration rates increases RMSE values steeply for estimated velocity and direction. However, lower penetration rates still provide RMSE values below \qty{5}{\meter\per\second}, even with vehicles obscured by building shadows excluded from the simulations. In conclusion, the results provide convincing evidence that connected vehicles' spatial and temporal distribution and characteristics significantly contribute to cloud motion estimation. This potential could further be improved by utilizing additional sensors, for example, from mobile base stations. Nonetheless, the results also show that a high penetration rate will be critical for accurate estimates. The final judgement will heavily depend on the actual (with post-processing achievable) accuracy of automotive light sensors for irradiance sensing purposes. This leads to the central implication that further research should investigate how to use cheap (automotive) light sensors for solar irradiance sensing. 

\medskip
\textbf{Supporting Information} \par 
Supporting Information is available from the Wiley Online Library or from the author.

\medskip

\bibliographystyle{MSP}
\bibliography{solarrrl}

\end{document}